\documentclass[conference]{IEEEtran}
\usepackage{comment}
\usepackage{amssymb,mathtools}
\usepackage{nccmath}
\usepackage{adjustbox}
\usepackage{algorithm}
\usepackage{algorithmic}
\usepackage[algo2e]{algorithm2e}

\usepackage{graphicx}
\usepackage{subcaption}

\usepackage{circledsteps}

\usepackage{diagbox}
\usepackage{multirow}

\ifCLASSOPTIONcompsoc
  \usepackage[nocompress]{cite}
\else
  \usepackage{cite}
\fi
\ifCLASSINFOpdf
\else
\fi

\usepackage{graphicx}
\usepackage[table, svgnames]{xcolor}

\usepackage[english]{babel}
\usepackage[utf8]{inputenc}
\usepackage{fancyhdr}

\begin{document}

\title{Too Expensive to Attack: Enlarge the Attack Expense through Joint Defense at the Edge}

\author{\IEEEauthorblockN{Jianhua Li}
\IEEEauthorblockA{\textit{Deakin University}\\
Melbourne, Australia \\
jack.li@deakin.edu.au}
\and
\IEEEauthorblockN{Ximeng Liu}
\IEEEauthorblockA{\textit{Fuzhou University}\\
Fuzhou, China \\
snbnix@gmail.com}
\and
\IEEEauthorblockN{Jiong Jin}
\IEEEauthorblockA{\textit{Swinburne University of Technology}\\
Melbourne, Australia \\
jiongjin@swin.edu.au}
\and
\IEEEauthorblockN{Shui Yu}
\IEEEauthorblockA{\textit{University of Technology Sydney}\\
Sydney, Australia \\
shui.yu@uts.edu.au}
}

\maketitle

\begin{abstract}
The distributed denial of service (DDoS) attack is detrimental to businesses and individuals as people are heavily relying on the Internet. Due to remarkable profits, crackers favor DDoS as cybersecurity weapons to attack a victim. Even worse, edge servers are more vulnerable. Current solutions lack adequate consideration to the expense of attackers and inter-defender collaborations. Hence, we revisit the DDoS attack and defense, clarifying the advantages and disadvantages of both parties. We further propose a joint defense framework to defeat attackers by incurring a significant increment of required bots and enlarging attack expenses. The quantitative evaluation and experimental assessment showcase that such expense can surge up to thousands of times. The skyrocket of expenses leads to heavy loss to the cracker, which prevents further attacks. 
\end{abstract}



\section{Introduction}
\IEEEPARstart{T}{he} Internet of things (IoT) creates unprecedented volumes of data in our living and working contexts. It is neither realistic nor necessary to transmit all data to remote clouds due to bandwidth and latency limitations. Thanks to its proximity to data sources, edge/fog computing is rapidly emerging as a game-changing way of hosting services, caching data, and expediting IoT applications \cite{li2018virtualfog}. Quite often, edge servers have some form of connection to the cloud for management, monitoring, and running services. In contrast to cloud devices, such edge servers are more vulnerable to various cyber threats, while the distributed denial of service (DDoS) is among the most notorious.

\begin{figure}[hbt!]
\centering
\includegraphics[width=0.9\columnwidth]{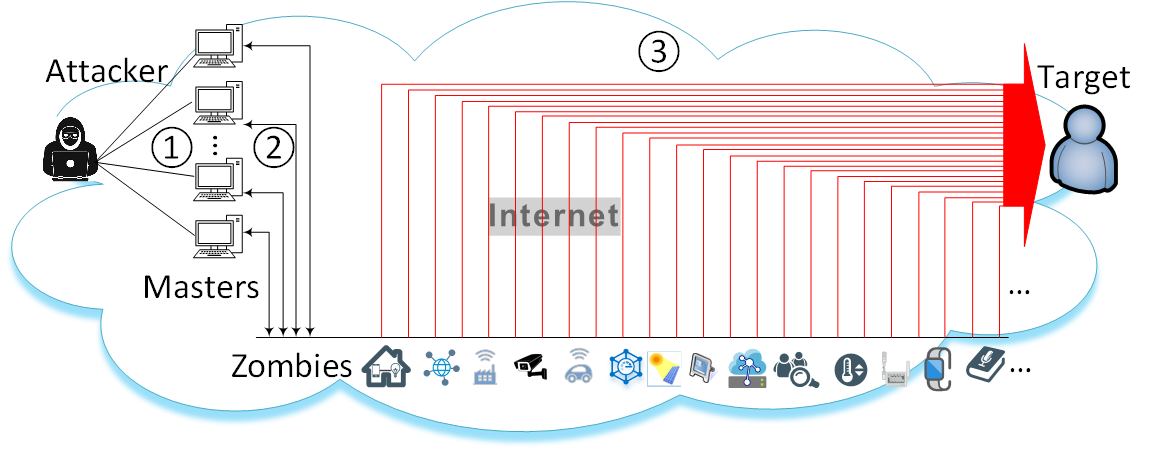}
\caption{Botnets-based attack: an attacker builds a group of masters (step \raisebox{.5pt}{\textcircled{\raisebox{-.9pt} {1}}}) from that to capture and control zombies (step \raisebox{.5pt}{\textcircled{\raisebox{-.9pt} {2}}}), then using distributed zombies to attack a user (step \raisebox{.5pt}{\textcircled{\raisebox{-.9pt} {3}}}). Such zombies are computing nodes locates in distributed networks. Due to a lack of collaboration between users, a user's network becomes an attacking source while another user's network is the victim.}
\label{f1}
\end{figure}

The primary problem with DDoS is that they can overwhelm various victims, such as servers, IoT devices, bandwidth, and even the Internet. Cybercriminals utilize readily available tools like a botnet-for-hire service to defeat businesses and demand payments \cite{li2020fleam}. As an essential element of botnets, a bot is a lightweight software instance running on a device automatically and autonomously. Such devices turn into networked zombies under the uniform control of an attacker, retrieving and carrying out instructions for illicit hacking purposes. As shown in Fig. 1, an army of zombies sends a stream of malicious packets with the target's IP address as the source IP to other uninfected nodes (known as reflectors), exhorting these machines to connect with the target. Then, these reflectors try to connect and send more traffics to the target. According to Amazon Web Services (AWS), the worst DDoS attacked bandwidth of more than 2.3 Tbps in 2020.\footnote{https://www.zdnet.com/article/aws-said-it-mitigated-a-2-3-tbps-ddos-attack-the-largest-ever} To this end, DDoS ransom attacks have flooded newspapers for money or Bitcoins \cite{mansfield2015growth}, while victims include airlines, banks, hospitals, local governments, to name a few. This new type of economical DDoS has become increasingly popular since the mid-2000s \cite{choi2010ddos}. Such attackers demand payment by preventing legitimate users from accessing rather than paralyzing entire networks and systems. 

Since an attacker coordinates zombies from every possible location, defenders should form alliances and collaborate in a distributed manner to fight with DDoS. Meanwhile, the detection is best done near the victim, while mitigation is most effective at the source end. Furthermore, an attacker cannot control the data transmission path, so defenders can add more checkpoints along the way to weaken the attack intensity. More interestingly, there is a lack of quantitative analysis on expenses at the attacker side. Therefore, our work fills in the gap, focusing on beating attacks at the edge. Our contribution is fourfold:


\begin{table}[t!]
\centering
\label{T1}
\caption{Symbols used in the paper}
\begin{adjustbox}{max width=0.95\columnwidth}
\begin{tabular}{|l|l|} 
\hline
Symbols                 & Description                                           \\ 
\hline
$\alpha_i$  & Positive constants signalling killing power           \\ 
\hline
$\epsilon$  & An accuracy factor used in estimation   \\ 
\hline
$\pi$  & Defense workloads allocation factor           \\ 
\hline
$C$                       & Capability matrix, $c_i$=1 can help, otherwise $c_i$=0      \\ 
\hline
$D_{ij}$                     & Defense capacity matrix                               \\ 
\hline
$d_{ij}$                     & User $i$ is capacity for protection of vulnerability j  \\ 
\hline
$f$                    & A series of functions                        \\ 
\hline
$\lambda_j$ & Times of defense units in the alliance                \\ 
\hline
$LA$                 & The value of local alarm                                    \\ 
\hline
$N_b$                 & Population of bots                                    \\ 
\hline
$N_d$                 & Number of defense units                               \\ 
\hline
$R$                       & Relationship between population of both parties       \\ 
\hline
$R^*$                      & The maximum value of R                                \\ 
\hline
$T$                      & The interval for routing discovery                                \\ 
\hline
$U_i$                      & User i                                                \\ 
\hline
$V_j$                      & Vulnerability j                                       \\
\hline
$X_{ti}$                      & The time to live of a route on the $i-$th link     \\
\hline
\end{tabular}
\end{adjustbox}
\end{table}

\begin{itemize}
    \item We revisit DDoS and pinpoint two problems, including inadequate consideration of the expense of attack and the collaboration of defenders. 
    \item We model the the population dynamics of bots and defense units in a combat, and prove that attack and defense are a game of resource competition. The increment of defense units leads to the growth of bots population, incurring a higher expense. 
    \item We develop a joint defense framework, allowing a defender to instantly acquire more defense units from collaborators via path manipulation. It offsets the scale advantage of the attacker. We overcome the constraints of path manipulation in terms of workload distribution, multiple path forwarding, and the loop-free condition. 
    \item  An experimental assessment and a quantitative evaluation are conducted to showcase the efficacy and the enlarged expense (up to thousands of times as previous). The soaring expenses will prevent cybercriminals from launching more DDoS attacks. Besides, we show the improvement of user experience.
\end{itemize}

It is worth noting that our scope is limited to profit-driven attackers. The road map of this paper is as follows. Section II studies related work, and Section III present the system model and assumptions in Section IV. We investigates the population dynamics during combat and incurred expense in Section IV, followed by our joint defense framework in Section V. Then, Section VI scrutinizes the proof of concept. In Section VII, we review the significance and conclude the paper. Table I shows symbols used for better readability.

\section{Related Work}
There are three locations to detect and mitigate DDoS traffic along the path, including source-end, victim-end, and in-network \cite{zargar2013survey}. Source-based mechanisms focus on proactively stop an attacker from generating DDoS codes at/near the source of attacks. But it is challenging to detect each source or filter attack flows accurately and timely. At the victim-end, defenders can closely monitor the victim and detect anomalies, but they may hardly respond to the attack before it reaches the victim. Hence, in-network solutions find their prosperity where networking devices perform packet filtering and firewalling in combating DDoS at the intermediate surface. However, network-based mechanisms usually lead to high storage and processing overheads on the routers that are less affordable at the edge. Recently, the concept of moving target defense (MTD) has emerged to thwart DDoS attacks. MTD tries to modify and control attack surfaces and increase the difficulty and uncertainty for attackers by manipulating system configurations (IP address, port number, etc.). However, constraints on IoT and edge devices \cite{li2015ehopes} may limit the concrete efficacy of MTD \cite{cho2019mtd}.

Next, we review frameworks with some form of expense factors considered. In \cite{zheng2018dynashield}, Zheng \textit{et al.} proposed the DynaShiled architecture using cryptocurrency PoW to challenge customer's capability to access a server. A customer must solve a hash-based puzzle generated by the defender to obtain a token. A customer with a valid certificate could access the victim's network. The authors showcased the reduced expense compared with some cloud providers. This solution may introduce a severe latency in the IoT environment \cite{li2017latency}.

In \cite{yu2014distribution}, Yu \textit{et al.} demonstrated using idle resources in a cloud to enhance server resilience. When a server is under attack, it can request more resources from a shared pool in the cloud. Although the cloud may have sufficient idle resources, the expense to a defender may be less affordable in handling large populations of botnets. 

Rashidi \textit{et al.} developed a collaborative solution in \cite{rashidi2016cofence}, where a defender could redirect excessive traffics to a neighbor for filtering. The neighbor got credits for the resource contribution. The more credits a member gets, the more help it receives when needed. The authors intended to leverage network function virtualization (NFV) to implement their framework. As NFV offers a cheaper way to replace expensive proprietary devices, the solution is cost-effective. The authors only performed qualitative analysis on the expense for defenders.  

\begin{figure}[t!]
\centering
\includegraphics[width=0.7\columnwidth]{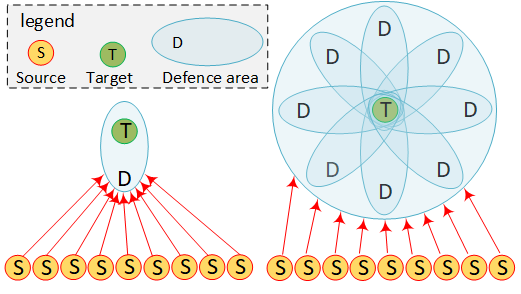}
\subfloat[\label{a} individual]{\hspace{.5\linewidth}}
\subfloat[\label{b} joint]{\hspace{.5\linewidth}}
\caption{The individual and joint defense.}
\label{f3}
\end{figure}

Thanks to the programmability, global view, and centralized management, SDN could remove the heavy reliance on other systems fighting against DDoS attacks. There are many DDoS framework proposed using SDN to provide an economical mechanism, including  \cite{bawany2017ddos} \cite{wang2015entropy} \cite{xiao2016efficient}. In particular, Bawany \textit{et al.} developed a DDoS framework on SDN in \cite{bawany2017ddos}, where the authors claimed their solution should be low-cost as SDN was an inexpensive platform for multiple services. Having said that, the authors did not provide evidence to prove the low-cost feature.

These lessons suggest that individual mitigation alone may not be sufficient to defeat DDoS attacks. In \cite{zargar2013survey}, Zargar \textit{et al.} argued that attacks with financial motivations were usually most dangerous and hard-to-stop. The authors concluded that botnets were the most dominant mechanisms launching DDoS attacks, where attackers cooperated to outwit a defender. Therefore, defenders should form alliances and collaborate in a distributed manner. Thus, we intend to offset the intensity of attacks on an individual defender, as depicted in Fig. 2. Next, we use the Lokta-Volterra Law to simulate the DDoS combat.

\section{System Model and Assumptions}
We assume that attack and defense are a game of resource competition. If an attacker has sufficient resources to incur more resource depletion to a defender than the defender can supply, the attacker will win, and vice versa.  The assumption is proved to be reasonable in Section IV. During an economic DDoS combat, a defender usually has a limited budget on security solutions. Let us assume a user ($U_i$) has a total number of defense units ($D_i$) within the budget. Generally, the investment in mitigation and defense should always be lower than the business gains received if there was no attack \cite{Somani2017cost}. Also, a cybercriminal has a budget and a reward for a successful attack. Cybercriminals will quit the combat when the expense is more than returns. We assume that the number of bots is proportional to the attack expense. 

Provided that the killing power of each unit of both parties is equal, i.e., one defense unit can handle one bot at a time, The defender must have at least one more unit to defeat the attacker, vice versa. In a collaborative model, one defender can instantly acquire more defense units from other collaborators. The increment of defense units in turn incurs the proportional growth of bots, leading to the surge of bot expense and the failure of an attack. The problem is how to add defense units during a combat. We solve this problem in our framework scrutinized in Section V.

A passive attack attempts to learn information from the system by eavesdropping and traffic analysis. It is hardly noticeable in contrast to an active attack that affects resources significantly. DDoS is an active attack in which cybercriminals prefer to use bots rooted in another user's network for stealth. IP spoofing \cite{zhang2016hadoop} is the process of modifying the source address to hide the identity of the sender and to impersonate another system. A defender can use the ingress filtering technique to examine incoming IP packets and use egress filtering to verify outgoing IP packets, ensuring the data with legitimate sources. It is the essential security requirement for a defender to become a collaborator, making it much more difficult for cybercriminals to launch DDoS from within the alliance. Hence, they are more likely to launch an attack from outside. Besides, we assume that defenders can share defense and attack information securely and timely. 

\section{The Interaction between the Attacker and the Defender}
Now, we investigate the interplay of resources and expenses.
\subsection{The Resource Competition}
During the combat, the specific number of active bots is up to the number of defense units $N_d$. The correlation obeys the Lokta-Volterra Law:

\begin{equation}
    \begin{cases}
       \frac{dN_d}{dt} &= \alpha_{1} N_d - \alpha_{2} N_d N_b   \\
       \frac{dN_b}{dt} &= \alpha_{4} N_d N_b - \alpha_{3} N_b 
    \end{cases}
\end{equation}
where $\frac{dN_d}{dt}$ and $\frac{dN_b}{dt}$ represent the instantaneous growth of the two populations, $t$ is time, $\alpha_{1}, \alpha_{2}, \alpha_{3}, \alpha_{4}$ are positive constants describing the interaction of the killing power of both parties. 

When neither of the population levels is changing, i.e., 
\begin{equation}
 dN_d = 0,  dN_b = 0 
\end{equation}

Combining Equation (1) and (2), we have

\begin{equation}
    \begin{cases}
       (\alpha_{2} N_b - \alpha_{1})N_d &= 0   \\
       (\alpha_{4} N_d - \alpha_{3}) N_b &= 0 
    \end{cases}
\end{equation}

The above equation yields two solutions:
\begin{equation}
       N_d = 0,  N_b = 0 
\end{equation}
and
\begin{equation}
 N_d = \frac{\alpha_{3}}{\alpha_{4}}, N_b = \frac{\alpha_{1}}{\alpha_{2}}
\end{equation}

Equation (4) tells the extinction of both populations. It is an invalid solution unless there is no attacker or defender in the world. Equation (5) represents a fixed level at which both parties sustain their non-zero levels. The two populations oscillate around the rooted value. 
\begin{equation}
 \frac{N_b}{N_d} = \frac{\alpha_{1}\alpha_{4}}{\alpha_{2}\alpha_{3}}
\end{equation}

We use the value of the constant of motion $R$ to represent the closed orbits approximately.

\begin{equation}
    R = N_b^{\alpha_{1}} e^{-\alpha_{2} N_b} N_d^{\alpha_{3}} e^{-\alpha_{4} N_d}
\end{equation}
The largest value of $R$ is thus attained that meets Equation (6):
\begin{equation}
    R^* = (\frac{\alpha_{1}}{\alpha_{2} e})^{\alpha_{1}}(\frac{\alpha_{3}}{\alpha_{4} e})^{\alpha_{3}}
\end{equation}
where $R^*$ is the maximum value of the constant.

\subsection{Attack Expense}
A cybercriminal leases botnets to carry out DDoS attacks on businesses and individuals. During an attack, the total resource consumed by benign and malicious data is:

\begin{equation}
    Resource_{attack} = Resource_{benign} + Resource_{botnet}
\end{equation}

From an attacker's perspective, benign traffic is free of charge, while there is an expense for the bot traffic. Cybercriminals may have other expenses on scanning, detection, and planning before an attack. To simply our modeling, we only count the botnet expense. Usually, the attacker needs to pay the expense for botnet setup and rental. The setup fee is charged once per attack, and the rental is payable for a lease period.

\begin{equation}
    Expense_{botnet} = Expense_{setup} + Expense_{rental}
\end{equation}

The setup expense per bot is relatively a fixed value per attack. The rental expense varies according to the population of a botnet and the duration. Renting around 50,000 bots costs between \$3,000-\$4,000 for 2 weeks.\footnote{https://www.bleepingcomputer.com/news/security/you-can-now-rent-a-mirai-botnet-of-400-000-bots/} If a bot is identified and mitigated, the attacker will lose the offensive power to the defender, so the mitigation response time (MRT) is a determining factor to the expense per bot during combat. To illustrate, if the MRT is more than two weeks, the per bot expense is only \$0.06 to 0.08 for two weeks. If the MRT is one hour, the per active bot expense ($PABE$) is \$20.16 to \$26.88 for two weeks.

The attacker has to incur sufficient attacking traffic to suppress resource supply for a given time. In consequence, the attacker must keep a certain number of active bots at the expense of:

\begin{equation}
    Expense_{botnet} = N_b \cdot PABE
\end{equation}
where $N_b$ is the population of bots.

\subsection{The Increment of Defense Units}
Let $D_{ij}$ represent the number of defense units belonging to user $U_i$ for the protection of vulnerability $V_j$.

\begin{equation}
D_{ij} = 
\begin{pmatrix}
d_{11} & d_{12} & \cdots & d_{1n} \\
d_{21} & d_{22} & \cdots & d_{2n} \\
\vdots  & \vdots  & \ddots & \vdots  \\
d_{m1} & d_{m2} & \cdots & d_{mn} 
\end{pmatrix}
(m,n \in \mathbb{N})
\end{equation}

Joint defense collaborators contribute to the protection for a victim. During a combat, the victim $U_k$ sends alarms to request possible assistance. Whether a collaborator $U_i$ can help is up to the $c_{i}$, $ \forall d_{ij} (i\neq k)$, $c_{i} = 1$ indicates that the collaborator can help, otherwise $c_{i} = 0$. 

\begin{equation}
C = 
\begin{pmatrix}
c_{1} & c_{2} & \cdots & c_{m}
\end{pmatrix}
(m \in \mathbb{N})
\end{equation} 
where $C$ is the capability matrix. The upstream neighbor has the potential to protect the downstream ones, i.e., $c_i=1$.

The joint defense power against the attack on vulnerability $V_j$ is 
\begin{equation}
    f(j) = \sum_{i=1,j=1}^{m,n} c_i \cdot d_{ij} \,\, (c_i \in [0,1])
\end{equation}

\begin{equation}
  \begin{split}
     \max \,\, f(j) &= \max \sum_{i=1,j=1}^{m,n} c_i \cdot d_{ij} = \sum_{j=1}^n  \Vert d_j \Vert \\
        &s.t.\left\{\begin{array}{lc}
        \sum_{j=1}^n d_{ij} \leq \Vert D_{ij} \Vert \\
        (c_i \in [0,1]) \end{array}\right.
  \end{split}  
\end{equation}
where $\Vert D_{ij} \Vert$ is the general defense power of all collaborators participating in the joint defense framework, $\Vert d_j \Vert$ is the joint defense power for the protection of $V_j$.

Thanks to the collaboration, a defender can have $\lambda$ times of troops than as it has in individual defense. 
\begin{equation}
  \begin{split} 
    \lambda_j &= \frac{\sum_{i=1,j=1}^{m,n} c_i \cdot d_{ij}}{d_{ij}} \gg 1 \\
        &s.t.\left\{\begin{array}{lc}    
    \sum_{i=1}^m c_i > 1 (c_i \in [0,1])\\
    0\le i \le m \in  \mathbb{N}\end{array}\right.
  \end{split}         
\end{equation}

According to Equation (6), the cybercriminal now needs $\lambda_j$ times of bots as much as the previous attack. In this case, the attack expense is at least  $\lambda_j$ times according to Equation (11).

\section{Joint Defense Framework}
We elaborate the joint defense framework as follows.
\subsection{Context and Benefits}
Collaborators include service providers (SP), businesses, and individuals. Note that our framework is complementary to, rather than replacing of any current solutions. The defense power is enhanced grounded on at least two facts. First, an attacker has little control over the route, and the malicious data travels along the shortest path towards the victim. Second, the surging growth of defense units along the route offsets the intensity of malicious traffics. 

\begin{figure}[t!]
\centering
\includegraphics[width=0.9\columnwidth]{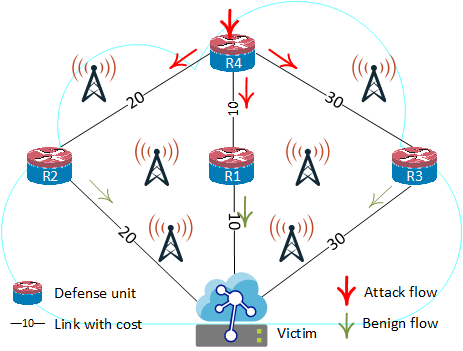}
\caption{The feasibility of joint defense: Data go through the shortest path by default (R4-R1-victim). When the victim is under attack, the upstream router R4 distributes part of data of low-priority to sub-optimal paths for mitigation.}
\label{f3} 
\end{figure}

When two agents swap high-level information directly, they form a peering relationship. Such a peering relationship reflects the closeness of a pair of upstream and downstream neighbors. The downstream neighbor has more accurate knowledge to judge benign or malicious codes as it is closer to a victim. However, the upstream neighbor has a better position to either prioritize or discard the data. The collaboration contributes to the accurate detection and recognition of bots at an earlier stage, leading to successful source-end mitigation. 

Fig. 3 further demonstrates such relationships between collaborators. We assume R4-R1 to be the shortest path from the cloud to the edge server such that the two routers maintain peering. As the victim, the edge server has the highest confidence of differentiating benign from the malicious. The victim notices the agent, and the agent sends an alarm to other agents asking for help. The alarm acknowledges the source of malicious codes with a combination of the source IP address and the victim's IP. Defenders can deliver part of the excessive data on sub-optimal paths such as R2 and R3. Based on the more accurate knowledge from the victim, all routers fight together with the attacker. At this moment, R4 has three peers of R1, R2, and R3, via which to send data to the victim. 

Note that R4 may request its upstream neighbor for help. In this regard, the upstream agents adjust the threshold on related defense nodes. Step by step, the collaborator and agents get to know the attack source. Next, the alliance can perform mitigation near the source-end, proactively protecting digital assets within the domain. 
 
\subsection{The Framework}

\begin{figure}[t!]
\centering
\includegraphics[width=0.8\columnwidth]{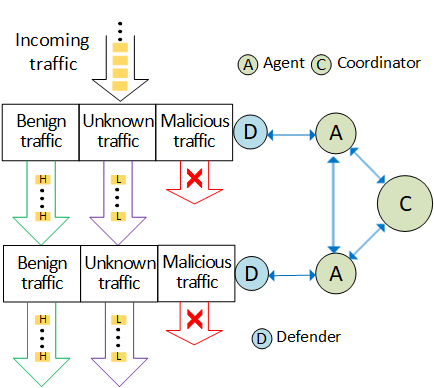}
\caption{The joint-defense framework: defenders perform marking, filtering, rate limiting to incoming traffic as usual. Agents monitor defense units, modify settings, and adjust thresholds. Agents and coordinators send queries, updates, and alarms to facilitate collaboration.}
\label{f4} 
\end{figure}

As illustrated in Fig. 4, the alliance has a coordinator-agent system to facilitate the cross-defender collaboration. The coordinator invites potential defenders to join the association who meet the above security requirement. Each defender installs an agent with a valid certificate issued via human channels. The primary responsibility of the agent is to share high-level information. Such information includes membership, identified bots, legitimacy test results, and knowledge of botnets. Agents swap updates and send an alarm to request helps. Meanwhile, agents monitor defense units, modify settings, and adjust thresholds in defense nodes. In more detail, a local alarm ($LA$=1) indicates the emergency within a zone of local businesses and individuals, while a regional alarm ($RA$=1) indicates the mess at the SP level.

\begin{figure}[t!]
\centering
\includegraphics[width=1\columnwidth]{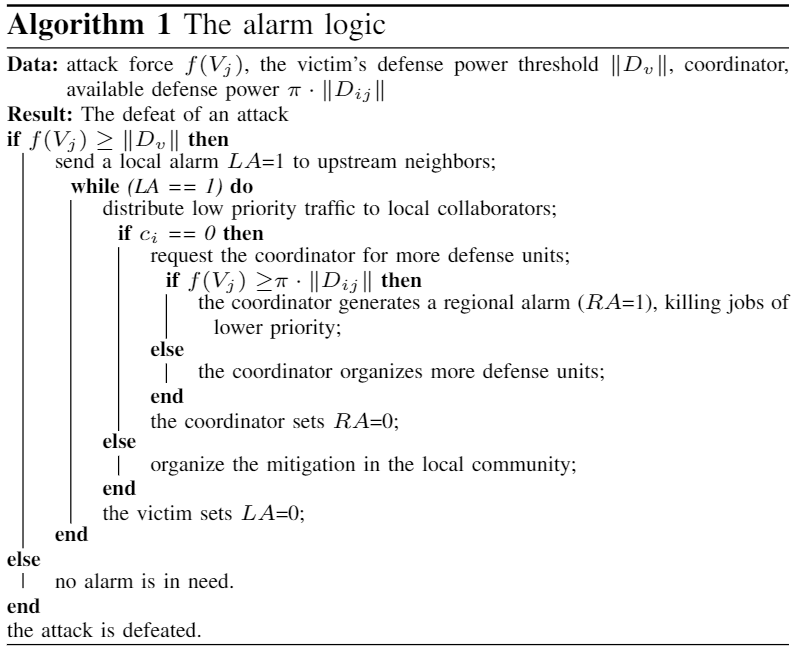}
\label{f4} 
\end{figure}

Algorithm 1 shows the alarm logic during the combat. In brief, the alarm mechanism is to prevent a network from being overwhelmed by excessive attacks. If the attack power is higher than the threshold, the agent raises a local alarm and sends a request to peers for possible offloading. The local alarm remains until the attack is defeated. If the victim, together with the requested peers, cannot suppress the attack, they request the coordinator to allocate more defense units. If the attack power is beyond the overall capacity, the coordinator and agents guide defense nodes to kill jobs of lower priority.

With a global view of the defense alliance, the coordinator has the duty for reputation evaluation, membership development, management, policy, etc. The coordinator propagates alarm messages to each agent and informs each agent of actionable knowledge of potential bots. At the same time, each defender employs individual infrastructure to take practical actions to combat DDoS. Grounded in guidelines and updates retrieved from the coordinator, each participant can prevent, detect, and mitigate bots in collateral networks. 

A defender keeps the current defense infrastructure running as usual. The classifier categorizes incoming data into benign traffic, marked with a higher priority `H'; data of uncertainty, marked with a lower priority `L'; and malicious data to drop. Initially, we proposed higher-priority and lower-priority marking, as the classification capacity of individuals may be limited. When two peers support more classes of traffic separation, they could implement more granular traffic classification. For example, the `H' refers to 100\% legitimacy, while businesses can have 95\%, 85\%, 75\% levels of legitimacy to reduce the risks of fault classifications. Without such granularity, the fault (i.e., malicious data marked as benign ones) could only be identified and processed at the victim's end. 

The framework has a different policy to process such data, including bandwidth reservation, reshaping, rate limitation, drop eligibility marking, etc. The purpose is to guarantee the Quality of Service (QoS) for benign traffic of legitimate users \cite{li2021scheme}. Meanwhile, defenders reserve enough resources for data of uncertainty detected by nodes of more accurate knowledge near the victim. The victim determines the attributes and informs other defenders. In case of excessive data entering a server suddenly, the local agent sends an alarm to other defenders and the collaborator. Then, they manipulate the data path, forcing such traffic to go through more defense units.

\subsection{The Constraints Analysis}
There are two constraints for collaborators in processing such traffics, i.e., load and multiple path routing. The collaborator must carry out offloaded jobs along multiple paths without incurring a routing loop. 

\subsubsection{The Workloads}
We formulate the following linear programming problem to minimize the maximum defense workloads \cite{he2014evolving}.

\begin{equation}
  \begin{split}
  &\min \,\, \pi\\
  &s.t.\quad \left\{\begin{array}{lc}
      \sum_{j=1}^n f(V_j) \leq \pi \cdot \Vert D_{ij} \Vert\\
      \sum f({U_i,V_j)} \leq \Vert D_i \Vert  \\
      vars. \,\, f(V_j) \geq 0, \forall  (U_i  \in  U, V_j  \in V)
      \end{array}\right.
  \end{split}
\end{equation}
where $f(V_j)$ is general attacking power for a cybercriminal against a vulnerability $V_j$, scaled by the defense workloads factor $\pi$, and $f({U_i,V_j)}$ is the job allocated to user $U_i$. The condition implies the workload capacity limitation for the alliance as well as the demand for a collaborator.

\begin{equation}
    \pi_i \triangleq \pi_i (1 + \epsilon \frac{\sum f({U_i,V_j)}}{\Vert D_i \Vert}
\end{equation}
where $\epsilon$ is an accuracy factor.

\subsubsection{Multiple Path Forwarding}
An upstream collaborator distributes offloaded traffic along multiple paths embedded with collaborative defense units. We assume that there are $M$ links towards the victim. $X_{tm}$ signals the time to live of a route on the $m$-th link following an independent distribution. We adopted the multiple paths routing numbers of $M$, with a random variable $T$ denoting the interval for routing discovery.
\begin{equation}
    T = max(X_{t1}, X_{t2}, \cdots, X_{tm})
\end{equation}
The cumulative distribution function of $T$ can be defined as follows:
\begin{equation}
    f_T(t) = P\{T \leq t\} = \prod_{i=1}^M f_{X_{ti}}(t)
\end{equation}

\begin{equation}
    f_{X_{ti}}(t) = 1 - \delta_i e^{-\delta_{i}t}
\end{equation}
where $\delta_i e^{-\delta_{i}t}$ is the probability density function of $X_{ti}$. Combining Equations (20) and (21), 
\begin{equation}
    f_T(t) = \sum_{i=1}^M \delta_i e^{-\delta_{i}t} \prod_{j=1}^{i-1}(1 - \delta_i e^{-\delta_{i}t}) f_{X_{ti}}(t) \prod_{j=i+1}^M(1 - \delta_i e^{-\delta_{i}t})
\end{equation}

To avoid routing loops, we set up a loop-free condition for data forwarding. We denote the shortest distance between an upstream neighbor and the victim as $f(u,v)$. The distance between users $U_i$ and $U_j$ is $f(U_i,U_j)$, while the distance from $U_j$ to the victim is $f(U_j,v)$. The user $U_i$ offloads the job to $U_j$ only when
\begin{equation}
    f(U_j,v) < f(U_i,v)
\end{equation}
Next, we showcase the efficacy of our framework and conduct the expense evaluation.
\section{Proof of Concept}
With deployments of IoT devices and the arrival of 5G, ultra-dense networks represent the trend for running services at the edge. There will be up to 50 networks per square kilometer in 5G \cite{he2021game}. No matter where the malicious codes come from, such traffics must traverse a series of networks. We first study the efficacy of collaborators to offset the intensity of attacks. 

\begin{table}[t!]
\label{T2}
\centering
\caption{The selected device and module}
\begin{adjustbox}{max width=0.8\columnwidth}
\begin{tabular}{|l|l|l|} 
\hline
\multicolumn{1}{|c|}{Device Name} & \multicolumn{1}{c|}{Model Used} & \multicolumn{1}{c|}{Function}  \\ 
\hline
R1                                & ethernet4\_slip8\_gtwy          & A network behind the Internet  \\ 
\hline
R2                                & ethernet4\_slip8\_gtwy          & A victim network               \\ 
\hline
Col1, Col2, Col3               & ethernet2\_slip8\_firewall      & A group of collaborators       \\ 
\hline
Internet                          & ip32\_cloud                     & The Internet connections       \\ 
\hline
Bots                              & 100BaseT\_LAN                   & Malicious traffic source       \\ 
\hline
Client                            & ethernet\_wrstn                 & Benign traffic source          \\ 
\hline
Server                            & ethernet\_server                & The victim                     \\
\hline
\end{tabular}
\end{adjustbox}
\end{table}

\begin{figure}[t!]
\centering
\includegraphics[width=0.8\columnwidth]{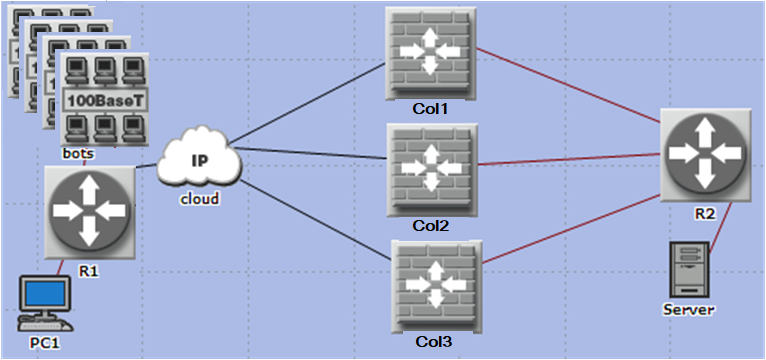}
\caption{The network diagram.}
\label{f5}
\end{figure}

\subsection{The Efficacy}
Imagine that an edge server manages access control and building security through IoT devices for a shopping mall. A cybercriminal demands payment for the ``protection". The cybercriminal will launch DDoS attacks on the server, causing the closure of the entire building. The attacker can either saturate the link between the edge server and the gateway or exhaust critical computing resources (CPU, memory, etc.) in the server. The management decides to join the alliance to fight the attacker. The appliance deploys a hybrid of protections, namely, victim-end, in-network, and source-end mitigation.

Fig. 5 illustrates the testbed implemented with the OPNET Modeler where there are three collaborators, namely, Col1, Col2, and Col3. Table II presents the models of the components in this experiment. On top of this, we use HTTP (image browsing) with the Page Interval Time of 0.03 seconds to represent the bot data and Database (high load) with Transaction Internal Time of 10 seconds to denote benign flows. In this setting, Col2 is the primary defender of the shopping mall. Col1 and Col3 are nearby collaborators. The DDoS flow may overstrain the computer power of Col2 when it performs mitigation individually. In this instant, the agent notices other collaborators for help. An upstream collaborator manipulates the data path, guiding the traffic to the victim in a load-balancing manner. Then, Col1 and Col3 join the game. We observe the CPU utilization at the primary defender during combat that lasts for about 2 hours. We suppose that the cybercriminal launches a 5-minute attack every 15 minutes.

We harvest the data when various numbers of defenders participating in the combat and present the result in Fig. 6. As studied in Section II, in-network mitigation may trigger a high overhead on routers. Note baseline stands for no DDoS attack in this evaluation. Fig. 6 (a) investigates the defender's load at the victim end. The CPU is nearly exhausted when it fights with the bots alone. When other collaborators participate in the combat, the CPU utilization drops significantly. Namely, the other collaborators offload the overhead in the alliance. 

\begin{figure}[t!]
\centering
\includegraphics[width=0.7\columnwidth]{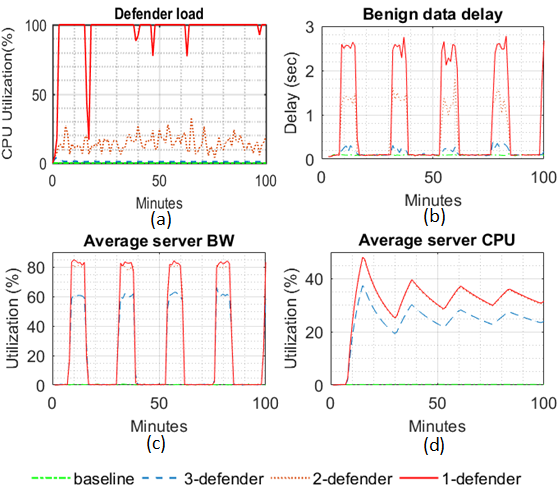}
\caption{The evaluation: (a) CPU utilization in the primary firewall, (b) benign data latency, (c) average bandwidth utilization between server and gateway, (d) average server CPU utilization.}
\label{f6}
\end{figure}

Fig. 6 (b) compares the benign traffic delay of legitimate users. Initially, the baseline is 0.09 seconds on average. During the attack, it goes up to 0.86 seconds on average. It is up to 2.86 seconds at the worst case in the 1-defender scenario. The delay drops to 0.48 and 0.14 seconds in the 2-defender and 3-defender scenarios, respectively. Therefore, the joint-defense providers a better user experience.

Fig. 6 (c) and (d) show the resource depletion in the victim-end. Each attack pushes high the bandwidth occupation of the link between the victim and the gateway. In detail, it reaches above 85\% when the defender fights alone on average. By offloading to other collaborators, it drops to about 60\%. Likely, the CPU utilization rate in the victim drops as more defenders work together. 

Thus, the alliance can handle the attack more efficiently in terms of resource consumption and service quality. Next, we explore the incurred expense of the attacker during the combat.

\subsection{Expense Evaluation}
It is difficult, if not impossible, to calculate how much money a cybercriminal needs to spend on botnets. Mirai botnet providers charge their users according to the number of bots (more bots, more money) and the attack duration (longer = more money). The least population is 1000 bots per rental for a minimum period of two weeks. In our joint defense, defenders enlarge their defense power against one attacker by placing more defense units. If the killing power remains constant, the growing number of bots is proportional to the increase of defense units for a successful attack. Therefore, We can roughly estimate how many times the expense of attack has increased from an individual mitigation solution. 

From Equation (11), a defender could incur a $\lambda_j$ times of expense than individual mitigation. The distance between two hosts on the Internet is about 16 hops \cite{paxson1997end}, so the middle point is 8. We limit our scope to ten primary DDoS threats (bandwidth, CPU, etc.). Then, we create a matrix $D_{8,10}$ with a random value between 1 and 100. Now, we get the $\lambda_j$ and present it in Table III.

A defender may have a certain amount of defense power for the protection of concrete vulnerability. Theoretically, every defender benefits from the collaboration regardless of its capacity. Technically, the less-powerful networks can get more gains from the alliance. For example, the defender $D_2$ incurs 3,977 times of expense on $C_6$. However, it is better to have a friend than an adversary in any fight. More collaborators mean less space for cybercriminals.

\begin{table}[t!]
\caption{The attack expense comparison}
\centering
\label{T3}
\begin{adjustbox}{max width=0.75\columnwidth}
\begin{tabular}{|c|c|c|c|c|c|c|c|c|c|c|}
\hline
   & C1  & C2  & C3  & C4  & C5   & C6   & C7   & C8  & C9   & C10 \\ \hline
D1 & 107 & 398 & 50  & 80  & 81   & 43   & 75   & 44  & 71   & 994 \\ \hline
D2 & 331 & 51  & 54  & 45  & 53   & 3977 & 110  & 52  & 97   & 92  \\ \hline
D3 & 46  & 166 & 57  & 306 & 45   & 71   & 40   & 199 & 166  & 362 \\ \hline
D4 & 41  & 221 & 58  & 72  & 55   & 209  & 47   & 181 & 137  & 56  \\ \hline
D5 & 44  & 92  & 663 & 147 & 71   & 41   & 41   & 51  & 58   & 52  \\ \hline
D6 & 44  & 61  & 64  & 994 & 3977 & 121  & 1326 & 86  & 62   & 67  \\ \hline
D7 & 284 & 65  & 95  & 110 & 110  & 249  & 568  & 166 & 1326 & 60  \\ \hline
D8 & 284 & 90  & 48  & 249 & 56   & 41   & 65   & 265 & 76   & 54  \\ \hline
\end{tabular}
\end{adjustbox}
\end{table}

In the market, Kaspersky Lab's experts reported that the average profit is about 2.6 times as of the investment in DDoS attacks.\footnote{https://securelist.com/the-cost-of-launching-a-ddos-attack/77784/} It is more difficult to predict the time and duration of an attack as it is at a cybercriminal's will. Our framework kills the profit margin as the expense is much higher than the estimated profit. As a result, profit-driven cybercriminals will quit the combat and stop attacking our digital assets.

\section{Further Discussion and Conclusion}
Next, we discuss the significance and conclude this paper.
\subsection{Further Discussion}
The DDoS attack is one of the most detrimental tools for crackers as it triggers catastrophic loss to a victim. With this tool, cybercriminals make easy money for decades. The remarkable profit keeps propelling the scale of related DDoS attacks, as presented in Table IV. In the first quarter of 2021, there are about 22.4 global DDoS attacks every minute on average. An attacker benefits from widespread bots across the world, destroying each of the defenders at a time. Hence, we advocate defenders to collaborate in a distributed manner, thus reducing the advantage of attackers. The main drawback of the current solution is that they lack enough consideration for the expense of attackers.  

There is no generic solution generally applicable to put an end to the economical DDoS. Many papers discussed the expense of mitigation on the victim's side. The proposed technologies include cloud \cite{yu2014distribution}, NFV \cite{rashidi2016cofence}, and SDN \cite{bawany2017ddos}, to name a few. Many authors claimed that their solutions were at low cost because the underlying platform (e.g., SDN) was inexpensive. However, there is no direct evidence to support such a claim.   

In sharp contrast to their counterpart in a cloud, edge servers are more vulnerable to DDoS attacks. The attack source tends to be from the cloud or zombies rooted in other networks. Meanwhile, cutting-edge technologies like 5G are making edge networks much denser. Thus, it is necessary and practical to establish a defense alliance to protect digital assets at the edge. To this end, our work timely steps in addressing concerns raised. Overall, we provide a simple, easy to implement, while effective collaborative defense framework for businesses, individuals, and service providers.

\begin{table}[t!]
\caption{Reported DDoS attacks}
\centering
\label{T4}
\begin{adjustbox}{max width=0.8\columnwidth}
\begin{tabular}{|c|c|c|c|} 
\hline
Time      & Attack Scale                                                               & Reporter                                                & Main Methods                                                                     \\ 
\hline
Q1-2021   & \begin{tabular}[c]{@{}c@{}}2.9 million DDoS\\attacks reported\end{tabular} & \begin{tabular}[c]{@{}c@{}}NETSCOUT\\ASERT\end{tabular} & \begin{tabular}[c]{@{}c@{}}Mainly Application \\Layer~DDoS attacks\end{tabular}  \\ 
\hline
15-Feb-20 & 2.3 Tbps                                                                   & AWS                                                     & UDP flooding                                                                     \\ 
\hline
Jan-19    & 396 Gbps                                                                   & Imperva                                                 & SYNC flooding                                                                    \\ 
\hline
Mar-18    & 1.7 Tbps                                                                   & NETSCOUT                                                & \begin{tabular}[c]{@{}c@{}}Memcached~\\amplification\end{tabular}                \\ 
\hline
28-Feb-18 & 1.35 Tbps                                                                  & GitHub                                                  & Git attack                                                                       \\ 
\hline
Oct-16    & 1.2 Tbps                                                                   & Dyn                                                     & DNS flooding                                                                     \\ 
\hline
Sep-16    & 1 Tbps                                                                     & Scmedia                                                 & TCP Syn, Ack                                                                     \\ 
\hline
Jul-15    &                                                                            & GitHub                                                  & HTTP flooding                                                                    \\ 
\hline
Nov-14    & 500 Gbps                                                                   & Forbes                                                  &                                                                                  \\ 
\hline
Feb-14    & 400 Gbps                                                                   & Cloudflare                                              & NTP                                                                              \\ 
\hline
Jul-13    & 300 Gbps                                                                   & Cloudflare                                              & DNS flooding                                                                     \\
\hline
\end{tabular}
\end{adjustbox}
\end{table}

Overall, an attacker will never stop attacking as long as the profit is high enough in a commercial market. The high returns drive the arising of previous, current, and future attackers at every corner of the world. We advocate the joint defense framework for maximizing the benefits of closeness between participants in the community. Collaborators fight together to beat DDoS by decreasing the attacking profits gradually. 

\subsection{Conclusion}
Botnets enable cybercriminals to attack a victim at a low cost, rewarding them with enormous returns financially. An edge server in IoT is more vulnerable to DDoS attacks. We advocate joint defense where many defenders collaboratively combat with one attacker along the path towards the victim. During the combat, the coordinator and agents swap knowledge,  instruct defense units to defeat bots for the victim in a distributed manner. The victim has more accurate information while the upstream defenders have a better position for mitigation. More importantly, defenders can manipulate routes to the victim while an attacker cannot. Therefore, our framework is simple, effective, and easy to implement to beat DDoS and other distributed attacks. The user experience is much improved simultaneously. 

The joint approach forces an attacker to employ more botnet populations during an attack, incurring much higher expense and preventing profit-driven crackers. We furnish proofs of the incurred expense growing linearly with the increment of participants. In detail, the framework can enlarge the expenses up to thousands of times. The skyrocket of attack levy, in turn, effectively stops and stifles the attackers' attempts, ultimately defeats them, and protects our digital assets in the community. In the future, we will work on solutions to trace malicious codes sourced from outside of the alliance, increasing their exposures so that the criminals have nowhere to hide.

\section*{Acknowledgment}
This work was supported in part by the National Natural Science Foundation of China under Grant No. U1804263, and No. 62072109.
\vspace*{\fill}
\bibliographystyle{IEEEtran}
\bibliography{./fogddos.bib}

\end{document}